\documentclass{elsart}
\usepackage[xdvi]{graphicx}
\usepackage{amssymb}
\usepackage{natbib}

\begin{document}
\headsep0.25in

\begin{frontmatter}

To be published in ``Carbon-based magnetism: An overview of the magnetism
of metal free carbon-based compounds and materials", edited by T. Makarova
and  F. Palacio (Elsevier 2005) \vspace*{5cm}

\title{Induced Magnetic Order by Ion  Irradiation of Carbon-Based Structures}
\vspace*{-0.2cm}\author{P. Esquinazi},
\ead{esquin@physik.uni-leipzig.de}
\author{R. H\"ohne,}  \author{K.-H. Han\thanksref{pa},} \author{D. Spemann,}
\author{A. Setzer,} \author{M. Diaconu,} \author{H. Schmidt,} \author{and T. Butz}
\address{Institute for Experimental Physics II, University of Leipzig, Linn\'estrasse 5,
04103 Leipzig, Germany}

\thanks[pa]{Present address: Department of Physics,
Ume\aa~University, Ume\aa, S-90735 Sweden.}


\end{frontmatter}

\section{Motivation}
\label{motiv} Irradiation effects in graphite were one major
research area in the past, partially due to its application as a
moderator in thermal nuclear reactors.  Graphite is still a
material of choice for nuclear applications due to its low
cross-section for neutron absorption. The influence of irradiation
damage produced by different kinds of ions on several properties
of graphite was  reviewed by  \citet*{kelly} in chapter~7 of his
book. Recently, irradiation effects in carbon nanostructures were
reviewed by \citet*{banhart99}. The effect of neutron irradiation
on the magnetic properties of graphite has been studied in the
past and shows the expected results, i.e. the introduction of
lattice defects by irradiation produces a decrease in the
diamagnetism and an increase in the spin density \citep*{kelly}.
We are not aware of any study made in the past on the effects of
proton irradiation on the magnetic properties of graphite or
carbon-based structures.

When we started this research work on irradiation effects on the
magnetism of graphite we had two reasons to choose protons as
energetic particle. The first one is related to the analysis
method called PIXE (Particle Induced X-ray Emission) that uses
protons to get a map for all relevant impurity elements within a
sample depth of 30~$\mu$m for a proton energy of $\sim 2$~MeV in
carbon (see Sect.~\ref{impu} for details). A systematic and full
characterization of the magnetic impurity content in each of the
samples, and after each treatment or handling (it makes no sense
to start with a highly pure sample and then cut it with a steel
knife afterwards), is of primary importance and absolutely
necessary.

The second reason was based on early reports on room-temperature
ferromagnetic behavior in some carbon-based structures (see
references in \citet*{makareview}). From those early works our
attention was focused to the  magnetic properties found in
amorphous-like carbon prepared from different hydrogen-rich
starting materials where an increase of the saturation
magnetization with the hydrogen concentration in the starting
material was found \citep*{murata91,murata92}. The origin for the
magnetic ordering has been related to the mixture of carbon atoms
with sp$^2$ and sp$^3$ bonds, which was predicted to reach a
magnetization higher than in pure Fe \citep*{ovchi91}. Hydrogen,
on the other hand, was assumed to have a role only in the
formation of the amorphous carbon structure \citep*{murata92}. New
theoretical predictions, however, show that hydrogenated graphite
can display spontaneous magnetization coming from different
numbers of mono- and dihydrogenated carbon atoms
\citep*{kusakabe03}. Spontaneous magnetization may also appear in
the case of monohydrogenated zigzag edges \citep*{fujita} if the
distance between them is large or if they are not in parallel.

The advantage of proton irradiation is twofold: it enables us to
make an impurity analysis simultaneously to the implantation of
hydrogen. In this chapter we will review the main effects obtained
after proton irradiation in different carbon-based structures.
This chapter is organized as follows. In the next section we
provide the main characteristics of our irradiation facility. In
section~\ref{impu} we show an example of element analysis obtained
in one of the graphite samples used for the irradiation studies.
The irradiation effects are reviewed in section~\ref{eff}. This
section is divided in two main subsections that describe the
effects in oriented graphite and carbon-based thin films. In
section~\ref{ann} we discuss some of the effects observed after
annealing the sample at high temperatures in vacuum or after
leaving it at room temperature for a long period of time.

In this chapter we concentrate ourselves mainly on the effects
produced by proton irradiation in highly oriented pyrolitic
graphite (HOPG). Effects of irradiation with alpha particles will
be discussed shortly in section~\ref{alpha}. Iron and fluor
irradiation effects on HOPG and on diamond are currently under way
and the results will be published elsewhere in the future.

 \section{Irradiation Characteristics}
\label{ic} All the irradiations presented in this chapter were
done with the high-energy nanoprobe LIPSION of the University of
Leipzig. The accelerator is a single ended 3~MV SINGLETRON$^{\rm
TM}$ with an RF-ion source for protons and alpha particles. The
focusing system can deliver proton beams with diameters as low as
40~nm at very low currents of the order of 0.1~fA. For the present
irradiations we worked at 2.25~MeV and currents up to 500~pA in
the case of proton beam of diameter of $1-2~\mu$m (microbeam). We
have also irradiated samples with a broad beam of 0.8~mm diameter,
energy of 2~MeV and currents between $\sim 50$ and $\sim 150~$nA.
The X-ray detector is an Ortec HPGe IGLET-X and subtends 187~msrad
solid angle. The numerical simulations discussed in this section
are based on the code SRIM2003 \citep*{ziegler}. The samples were
mostly attached to a Si substrate using a small amount of Varnish
or a mixture of Varnish with ultra-pure graphite powder. In one
case we fixed the sample on a messing plate with a hole to enable
the free penetration of the energetic particles (H$^+$ and
He$^+$). The magnetic moment of the substrates was always measured
before attaching the sample. All the irradiations were done at
nominally room temperature; the temperature of the sample was not
controlled during irradiation.

\subsection{The interaction of MeV-protons with graphite}
 \label{irr}
\begin{figure}
\begin{center}
\includegraphics[width=\textwidth]{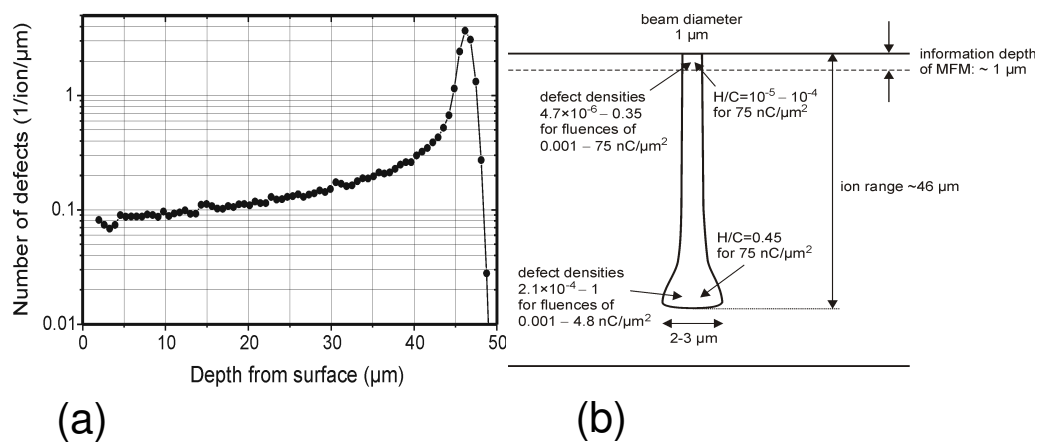}
\end{center}
\caption{SRIM2003 Monte Carlo simulations \protect\citep*{ziegler} for
2.25~MeV protons on HOPG. (a) The number of defects per ion and 1~$\mu$m
depth interval. The penetration range of 46~$\mu$m is clearly visible. (b)
Sketch of the modified area in graphite due to 2.25 MeV proton
bombardment. The lateral straggling has a full width at half maximum
(FWHM) of $\sim 2.5~\mu$m for an initial beam diameter of $1~\mu$m. The
numbers are obtained assuming a displacement energy of 35~eV for Frenkel
pairs in HOPG. Adapted from \protect\citet*{butz}.} \label{protons}
\end{figure}

Figure~\ref{protons}(a) shows a typical profile for the number of
defects created in a carbon matrix per 2.25 MeV-proton and per 1
~$\mu$m interval. The penetration range of about 46~$\mu$m is
clearly visible. For samples thicker than this range one expects
that the majority of the protons will be stopped, whereas for
samples substantially thinner essentially all protons traverse the
sample. This means that for $20~\mu$m or thinner samples the
implantation of protons should be undetectable. Strikingly,
enhancement of ferromagnetism has also been observed in disordered
carbon \citep*{hoh04} and fullerene films deposited on Si
substrates after proton irradiation (see section~\ref{thin}).
Further details on the different processes that influence the
penetration of protons in graphite as well as the restrictions of
the SRIM2003 simulations were discussed by \citet*{butz}. The
question how many protons remain in thinner samples cannot be
clearly answered yet. Experiments with thinner and unsupported
samples (to rule out part of the backscattering processes) are
necessary and will be performed in the future. Under the
assumption that there is no annealing of defects during
irradiation and that the damaged area is of the same size as the
irradiated one, the numerical simulations indicate that for
fluences $0.001\ldots 75~$nC/$\mu$m$^2$ we get in the near surface
region between $4.7 \times 10^{-6}$ and 0.35 displacements per
carbon atom, i.e. complete amorphization for the highest fluence,
using a displacement energy of 35~eV for Frenkel pairs in HOPG, in
agreement with recently published studies of the damage cascades
by irradiation on graphite \citep*{abe97}. Changing this number to
28~eV yields defect densities which are about 30\% higher. Towards
the end of the track, the lateral beam straggling becomes
important and the damaged area can be several microns wide, see
Fig.~\ref{protons}(b). For a fluence of $75~$nC/$\mu$m$^2$ we have
$\sim 5 \times 10^{11}$ protons/$\mu$m$^2$, i.e. the regions where
defects are created by each individual proton overlap heavily.

Note that even for the highest fluence the relative concentration
of hydrogen (H/C, see Fig.~\ref{protons}) in the first micrometer
from the surface remains rather low, of the order of $10^{-4}$ or
lower. However, most of what happens after the defect formation is
still unclear. A dangling bond could attract a hydrogen atom - not
necessarily from the proton implantation but already present as
impurity in the sample. Interestingly enough, there is little
information on residual hydrogen from pyrolysis of hydrocarbons,
possibly because detection methods for hydrogen contents of the
order of 100~ppm or below were not readily available and such
quantities were of little relevance for applications as moderators
in reactors. Hydrogen atoms in the van der Waals gap or within the
graphite layers should be highly mobile, contrary to hydrogen
atoms trapped at defects. Investigations with deuterium done by
\citet*{siegele} indicate that deuterium does not readily diffuse
out of HOPG but is rather chemically bound up to D/C ratios of
about 0.45, probably at lattice defects. The maximum retention of
deuterium depends on the temperature and implantation energy. A
broad range of binding energies for hydrogen in graphite up to
about 4~eV was reported by \citet*{atsumi}, i.e. even temperatures
as high as 3000-3500~K may be not sufficient to eliminate all
incorporated hydrogen. On the other hand, the effective activation
energies for hydrogen diffusion in our system are not necessarily
the same as, for example, those obtained by experimental methods
-- usually at high temperatures -- to study kinetics of diffusion
of hydrogen in graphite. Irradiation effects of MeV protons on
diamond-like films were studied by \citet*{wang92}. According to
the authors, below a fluence of the order of 1~nC/$\mu$m$^2$ the
hydrogen atoms produced by ion irradiation could be recaptured by
dangling bonds. For higher fluences a release of hydrogen is
expected.

Irradiation at a fixed energy has a clear disadvantage. If the
magnetic ordering is triggered at a specific density of protons
and defects, then, it is clear that with fixed proton and defect
distributions, such as the one shown in Fig.~\ref{protons}, we
would have a rather narrow window to get a maximum effect.
Therefore, it should not be a surprise if for some irradiation
energies and fluences one measures negligible effects or even a
reduction of the magnetic order present in the sample before or
after some irradiation steps. Carbon magnetism as well as
irradiation effects on it belong to a new field in magnetism. For
graphite bulk and thin film samples there are still many questions
to be clarified in the future concerning the hydrogen implantation
by irradiation and its effects on the magnetism.

The defect formation process by high energy protons is a non-equilibrium
athermal process and it appears rather unlikely that ordered arrays of
defects are formed by migration of interstitial carbon atoms or vacancies,
maybe with the exception of the interstitial across the gallery. According
to \citet*{banhart99} and from electron irradiation studies, the essential
types of radiation damage up to intermediate temperatures are the rupture
of basal planes (due to shift of the C-atoms out of the plane) and the
aggregation of interstitials into small dislocation loops between the
graphene layers. The migration energy of the interstitial depends whether
it is bounded. Di-interstitials were proposed to explain the
irradiation-induced amorphization of graphite with a migration energy of
0.86~eV \citep*{niwase95}. The interstitial loops are stable up to rather
high temperatures, probably to 1000$^\circ$C \citep*{banhart99}.
Certainly, the layer structure should remain essentially intact, at least
for our low fluences and at the sample surface.

\begin{figure}
  \begin{center}
  \includegraphics[width=12.5cm]{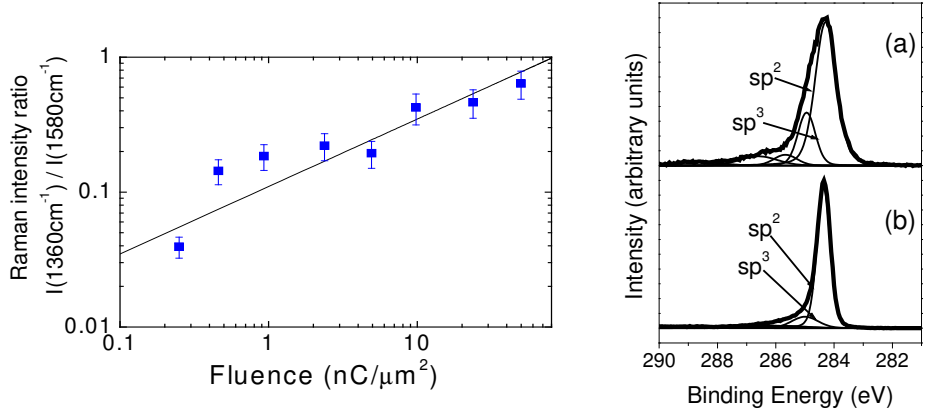}
  \end{center}
  \caption {Left figure: Raman intensity ratio as a function of the irradiation fluence
  for spots created on a HOPG sample (see section \protect\ref{spots}).
  The line represents the fit curve $0.1 x^{0.5}$.
  Right figure: Decomposition of the XPS C$_{1s}$ core-level peaks of HOPG surface
  after (a) and before proton irradiation (b). The components after peak fit are: at
  284.4 eV: ``C-sp$^2$"; at 285.0 eV: defect peak ``C-sp$^3$";
  other components: different C-O bonds.
  The thick and thin solid lines denote the experimental curves
  and the components, respectively. The measurement was done after irradiation
  of the sample with 1.2~mC and a fluence of 0.6~nC/mm$^2$ (step 4 of the
  sample reported by \protect\citet*{pabloprl03}). Adapted from \protect\citet*{hoh04}.}
 \label{raman}
\end{figure}

Irradiation can also produce a transition from sp$^2$ to sp$^3$ bonding
leading to cross-links between the graphene layers and the formation of
sp$^3$ clusters \citep*{tanabe96}. These clusters appear to be stable and
do not anneal at high temperatures. Our own Raman and X-ray Photoemission
Spectroscopy (XPS) measurements indicate the following. For all
unirradiated areas of the HOPG samples the Raman spectra show only one
pronounced peak at 1580~cm$^{-1}$ (the E$_2$g$_2$ mode) as expected for
graphite without any disorder. For the spots produced with the proton
microbeam the Raman spectra show additionally the disorder mode D at
1360~cm$^{-1}$. With increasing fluence the E$_2$g$_2$ and D modes become
broader and the ratio of their intensities I(D)/I(E$_2$g$_2$), as a
measure of the degree of disorder, increases (see Fig.~\ref{raman}).

The XPS results show a clear difference between an irradiated and
an unirradiated sample, see Fig.~\ref{raman}. The C$_{\rm 1s}$
core-level peaks of HOPG samples were recorded using a pass energy
of 10~eV and were fitted by six components with binding energies
of 284.4~eV (main peak of HOPG, C-sp$^2$), 285.0~eV (defect peak
``C-sp$^3$"), 285.8~eV, 286.3~eV, 289.0~eV (C-O components) and
290.9~eV (shake-up peak). The main result is that the defect peak
component increases after proton irradiation. A similar behaviour
was found for HOPG under plasma low-energy argon-ion bombardment
\citep*{rousseau03}. Whether this defect peak corresponds to a
pure C-sp$^3$ state or whether some of the ``defect" carbon atoms
are bounded to hydrogen is unclear and it is a matter of current
research \citep*{estrade04}. The knowledge of the C-H sp$^2$ and
sp$^3$ bonds is for the magnetic properties of carbon structures
of importance. We remark that numerical simulations indicate a
100\% polarized $\pi$-band, i.e. a ferromagnetic order stable at
room temperature, for a graphene layer with a mixture of sp$^2$
and sp$^3$ bondings (mono- and di-hydrogenated) at the zigzag
edges of a graphene layer \citep*{kusakabe03}.

Other aspect of the irradiation effects on graphite that may be of
importance to trigger magnetic ordering is the  formation of pentagons and
heptagons in the basal planes, which could cause a bending of the graphene
layers \citep*{banhart99}. The influence on the electronic band structure
of graphite of such topological defects has been studied theoretically by
\citet*{gon01}. According to these authors, these defects may trigger
ferromagnetism or even superconductivity.

For thick targets and sufficiently high fluences, the target surface
actually swells, which can be easily measured by an atomic force
microscope (AFM), see Fig.~\ref{height}(a), or is even visible under an
optical microscope \citep*{spemann04,pablopt}. The swelling in the
$c-$direction occurs together with the contraction in the graphene layer;
the new formed interstitial planes push the existing planes apart leading
to a protuberance at the sample surface. This irradiation effect was
studied by \citet*{koike94,muto97,muto97ii}. This effect was not observed
for thin enough targets for the usual proton irradiation fluences used in
this work.

\begin{figure}
  \begin{center}
  \includegraphics[width=12.5cm]{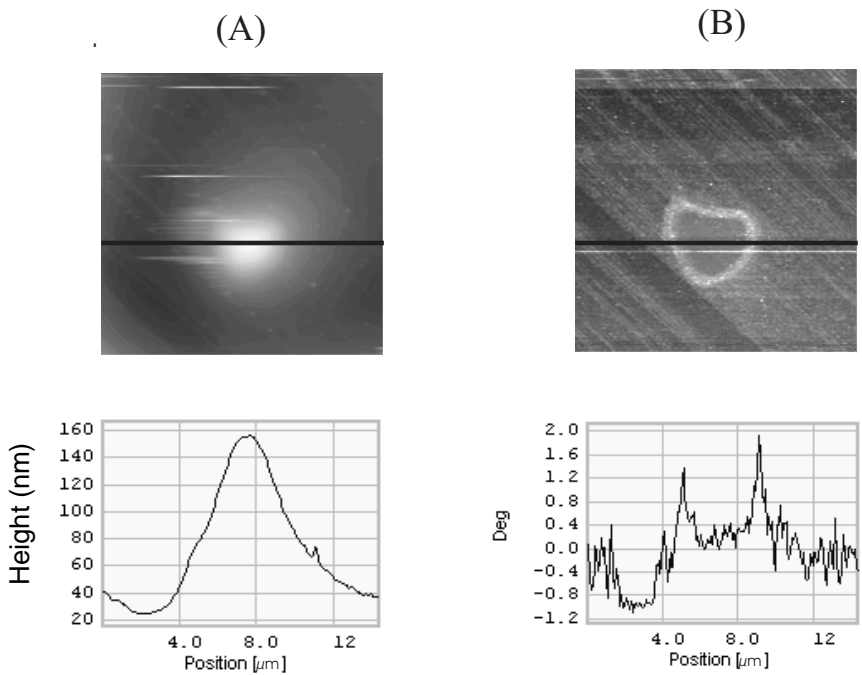}
  \end{center}
  \caption {(A) Topography and (B) MFM images obtained from a
spot produced with a dose of 6.8 nC/$\mu$m$^2$ and a current $I =
0.51~$nA, which means a current density larger than 5~mA/cm$^2$. The scan
area is $15~\mu$m$ \times 15\mu$m. The bottom figures show the line scans
done at the positions of the black lines of the upper pictures. The MFM
images were taken at a distance of 50~nm using standard magnetic tips.
Taken from \protect\citet*{pablopt}}
 \label{halo}
\end{figure}

Another important aspect regarding the irradiation effects is the
heat load due to the beam. For a spot of $2~\mu$m $\times 2~\mu$m
and a current of 500~pA this amounts to $\sim 300~$W/mm$^2$ for
thick samples. Most of the energy will be deposited near the end
of the ion track. Depending on the lateral dimensions and the
thickness of the sample as well as the target holder, heat is
transported away from the beam spot. The details are difficult to
calculate, but the measurement and control of the temperature is
of importance. For thin samples, this is relatively unimportant.
We have evidence that at higher currents the magnetic response in
MFM-measurements is strongest at the rim of the spot suggesting
annealing effects in the center of the spot, as Fig.~\ref{halo}
indicates. For lower currents the magnetic image is homogeneously
distributed across the spot (see section \ref{hopg}), in contrast
to the magnetic ``ring" shown in Fig.~\ref{halo}.

 \subsection{Impurity measurements}
 \label{impu}

\begin{figure}
\begin{center}
\includegraphics[width=\textwidth]{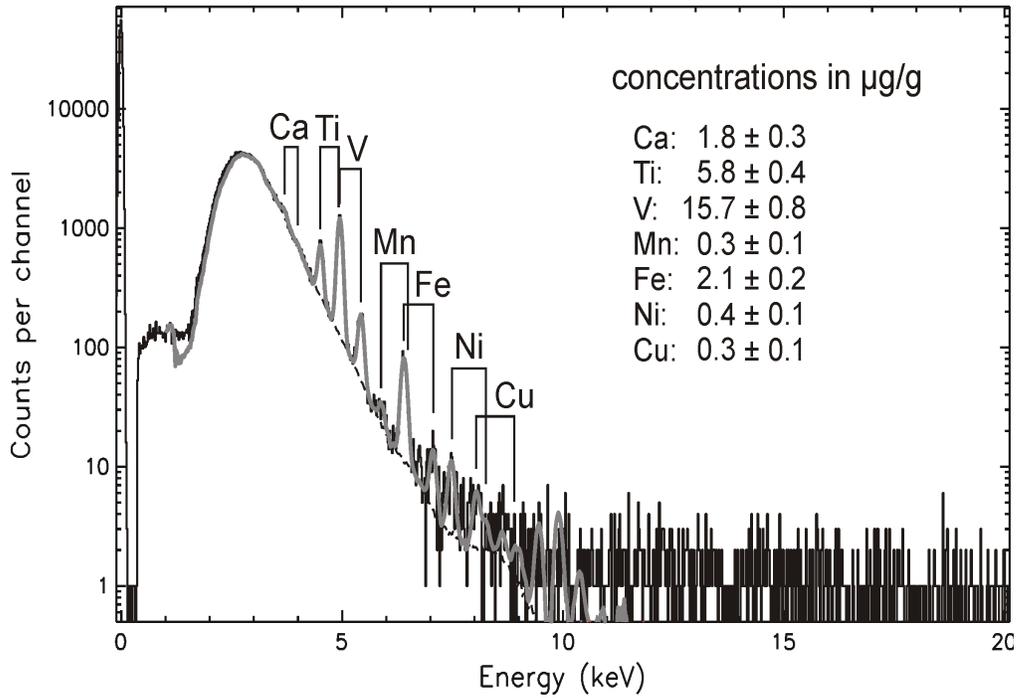}
\end{center}
\caption{Typical broad beam PIXE spectrum from a HOPG sample. The
main impurities are Ti, V, and Fe. The minimum detection limit for
other elements heavier than Si is $\sim 0.3~\mu$g/g. Note that
2.1~$\mu$g/g iron in the carbon matrix means a concentration of
0.45~ppm Fe. The background curve is due to the carbon matrix.}
\label{pixe}
\end{figure}
With the proton micro-beam a total charge of 0.5~$\mu$C suffices
to obtain a minimum detection limit for Fe impurities below
$1~\mu$g/g. With our broad-beam PIXE facility we work with 2~MeV
protons and currents of 150~nA and a beam diameter of 0.8~mm.
Here, about $5~\mu$C are required to reach a minimum detection
limit of $1~\mu$g/g for Fe. The fluences are in the range of  $1
\ldots 50~\mu$C/mm$^2$ for the micro-beam and the broad-beam,
respectively. These fluences are several orders of magnitude lower
than the fluences actually applied for the induction of magnetic
ordering in graphite, as will be discussed below. The advantage of
the micro-beam compared to the broad-beam is that a distribution
map for all relevant impurity elements can be obtained contrary to
the integral value for the broad-beam method, a rather important
issue considering the grossly inhomogeneous distribution of
impurities like Fe which we have encountered in some samples (see
for example Fig.~1 in \citet*{spemann03}). A typical broad beam
PIXE spectrum for a HOPG sample is shown in Fig.~\ref{pixe}. It
shows the presence of a number of impurities, the Fe content being
$(0.45 \pm 0.04)~$ppm. From the study of ferromagnetic signals in
different HOPG samples \citep*{pabloprb02} we may suspect that
HOPG could contain a non negligible amount of hydrogen, some of
them may be related to the origin of the magnetic signals. The
amount of hydrogen before or after irradiation within the fluence
used in this work cannot be determined by PIXE. Taking into
account recently done theoretical work done by \citet*{lehtinen04}
that estimates a magnetic moment of $\sim 1~\mu_B$ for a carbon
vacancy, $2.3~\mu_B$ in the surrounding of an hydrogen bonded to a
carbon atom at the position of a carbon vacancy, or $1.2~\mu_B$
for a vacancy that is saturated by two hydrogen atoms in a
graphene layer, we may expect that a few ppm hydrogen trigger a
non negligible magnetic signal in graphite. Therefore, hydrogen
content measurements should be carried out with a sensitivity in
the ppm range, a rather difficult task. The reader can read
\citet*{butz} for a short discussion on the methods available for
this kind of measurement.

 \section{Irradiation Effects}
 \label{eff}
\vspace{-0.5cm} \subsection{On highly oriented pyrolytic graphite}
 \label{hopg}
\vspace{-0.2cm}\subsubsection{Broad proton irradiation}

\begin{figure}[b]
\begin{center}
\includegraphics[width=\textwidth]{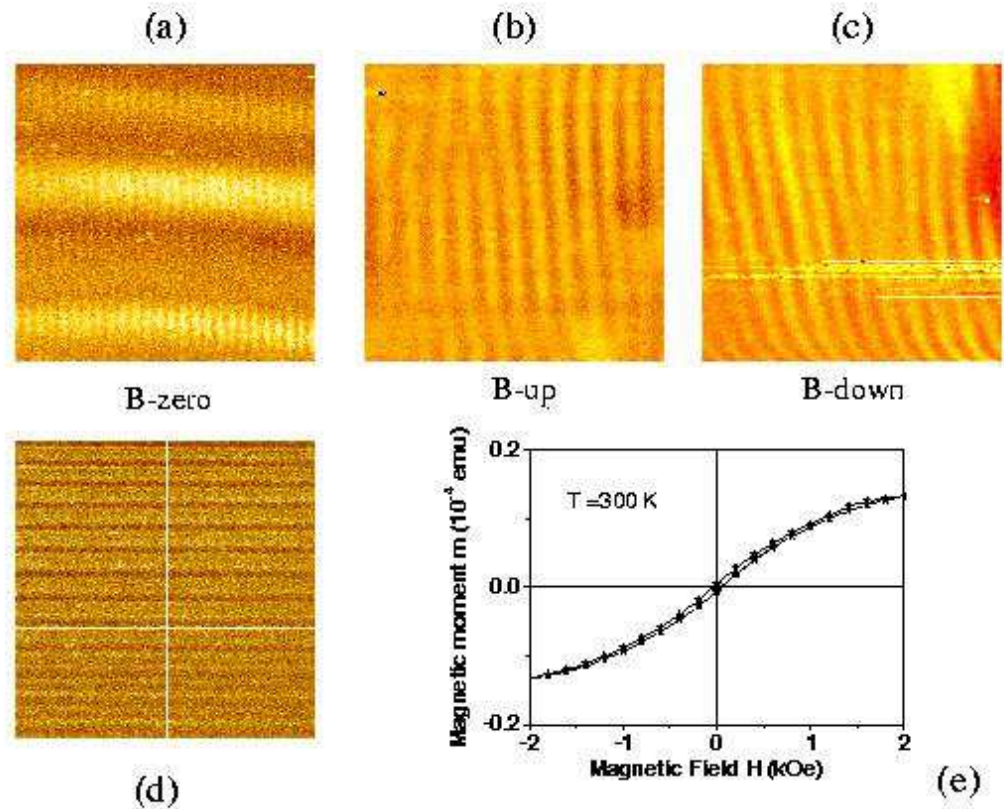}
\end{center}
\vspace{0.0cm} \caption{Magnetic force microscope (MFM) and SQUID
measurements of a HOPG sample irradiated with 4 broad spots ($\phi
\simeq 0.8$~mm) of 135~$\mu$C total charge each. All images scan
an area of $20 \times 20 \mu$m$^2$: (a) MFM image of the sample in
the state just after irradiation, (b) after application of a field
of + 1 kOe, (c) after application of a field of -1~kOe. All
measurements were done in remanent state after removing the field,
i.e. at no applied field. (d) Magnetic domain structure of the
sample in the same state as in (a) but at a different position.
All the MFM measurements were done with a standard magnetized MESP
tip and at a distance of 50~nm from the sample surface. (e) The
magnetic moment of the same sample measured with a SQUID at room
temperature. The sample was fixed on a Si substrate for the
irradiation. The data in the figure are the original data of the
sample with Si substrate, without any substraction. The signal of
the sample overwhelms that from the substrate. Measurements of the
sample standing alone reproduces basically the curve presented
here.} \label{reman}
\end{figure}
All the irradiated HOPG samples discussed in this review were from
Advanced Ceramics Co. (ZYA grade, $0.4^\circ$ rocking curve width
at half maximum) with a content of magnetic metallic impurities
below 1~ppm. The largest concentration of non magnetic metallic
ions was found for Ti ($\lesssim 6~\mu$g/g) and V ($\lesssim
16~\mu$g/g). A typical surface area of the samples was $2 \times
2~$mm$^2$ and a thickness between 0.1~mm to 0.3~mm. The samples
were  glued with varnish (or a mixture of varnish and a high
purity graphite powder to increase the thermal coupling) on a
high-purity Si substrate and the magnetic moment of the whole
ensemble as well as of the Si substrate alone was measured. In
general and within experimental error the diamagnetic hysteresis
loops for the used Si substrates are reversible; after subtracting
the linear diamagnetic dependence there is no significant
hysteresis left (for an example see Fig.~\ref{2I}(a)).
Nevertheless, when the ferromagnetic moment of the sample is weak,
it can be a hard task to obtain the true sample magnetic moment
from the SQUID signal of the mixture (Si plus the HOPG sample) and
several checks have to be done.

In this section we will discuss  results obtained after
irradiation of a large area of the HOPG sample using the micro-
and the broad-beam of protons. As we will realize below, there are
several irradiation parameters that may have an important role in
inducing the achieved
magnetic signal. Namely:\\
\noindent  -- (1) The total implanted charge. The total amount of protons
or alpha particles that were implanted in the sample or travelled through
the sample.\\
\noindent  -- (2) The input energy. Although with the LIPSION we
have the possibility of changing this energy from 1 to 2.25 MeV,
in order to minimize the number of variables, the studies
presented here were done with fixed irradiation energies of 2.0 or
2.25 MeV. This means that we have a well defined defect and
implantation profile inside the sample, see Fig.~\ref{protons}. If
the highest magnetic signal is determined at a given defect and
proton density, it is clear that successive irradiation at similar
energies may produce contrary effects if this optimized region
is destroyed by the next irradiation at similar conditions.\\
\noindent  -- (3) The fluence, the irradiated charge per unit area.\\
\noindent  -- (4) The ion current. Large currents might heat the sample
and non systematic effects are then possible.\\
\noindent  -- (5)  Micro- or macro-irradiation, i.e. a broad
irradiation with the 0.8~mm beam or a large number of micrometer
spots distributed in the sample. Experience indicates that the
magnetic signals are much larger when one uses a high density of
micrometer spots rather than the implantation of similar amount of
charge with a broad beam. This fact appears to be
related with the density and/or type of defects that the beam produces.\\
\noindent  -- (6) Sample temperature. All the irradiations
presented in this review were done at nominally room temperature.
Future experiments should try to irradiate at lower or higher
temperatures to check for its
influence.\\
\noindent -- (7) Finally, the initial state of the sample, namely the
density and type of defects, which in part can determine its metallicity.
Computer simulation results of the effects of adsorbed hydrogen on the
band structure of a graphene layer indicate that metallization caused by
specific defects can quench a spin polarized state \citep*{duplock04}.

The first SQUID measurements that indicate a magnetic ordering after
proton irradiation were published by \citet*{pabloprl03}. In that work an
increase in the hysteresis loop was observed after several irradiation
steps. These irradiation steps contained successively irradiation of
several thousands of micrometer small spots as well as four (or three)
spots of 0.8~mm diameter each on the same sample. In what follows we shall
discuss further work that has been done afterwards, where we have tried to
characterize the effects produced by some of the irradiation parameters
described above. We stress that they are the very first steps to get
reproducible magnetic order in carbon, a task that turned to be full of
difficulties, as the published results from literature indicate.

A broad and homogeneous proton irradiation of usual fluences
($\sim 150~\mu$C total charge in an area of $\sim 1~$mm$^2$) per
spot on oriented graphite produces a magnetic signal that in
general can be well observed with a magnetic force microscope
(MFM) on the irradiated surface. Figure~\ref{reman} shows the MFM
images obtained on an irradiated area with a total charge of
540~$\mu$C distributed into 4 spots, 135~$\mu$C each. After this
broad irradiation there is no significant change in the topography
worth to note. However, the periodic magnetic domains are well
defined. In the remanent state just after irradiation, in some of
the irradiated area of the sample we observed two domain
structures, one normal to the other, see Fig.~\ref{reman}(a). The
period of the small domain structure is $\sim 0.8 \ldots 1.2~\mu$m
depending on the region (compare with (d) which was obtained in
the same state but in other irradiated area), whereas the other
domain structure has a period of $\sim 10~\mu$m and depends on the
scan direction of the MFM tip. After application of a field of
1~kOe in the $+z$ direction (perpendicular to the graphene layers)
the ``vertical" domain structure shows a period of $\sim 1.6
\ldots 2.0~\mu$m and the width of the domains increases, see
Fig.~\ref{reman}(b). The other domain structure is not observed.
After magnetizing the sample with a field in the other direction
there is a slight change of the width of the bright relative to
the dark regions, compare (c) with (b). In non-irradiated graphite
areas we did not find any signature of domain structures within
the resolution of the microscope. This irradiation triggered a
relatively large hysteresis loop that could be very well measured
with the SQUID without any background substraction, see
Fig.~\ref{reman}(e). The temperature dependence of this signal was
studied removing  the sample from the Si substrate. It shows a
weak decrease ($\sim 5\%$)
 with temperature of the saturation magnetization and coercivity fields
between 5~K and 300~K.

\begin{figure}
\begin{center}
\includegraphics[width=\textwidth]{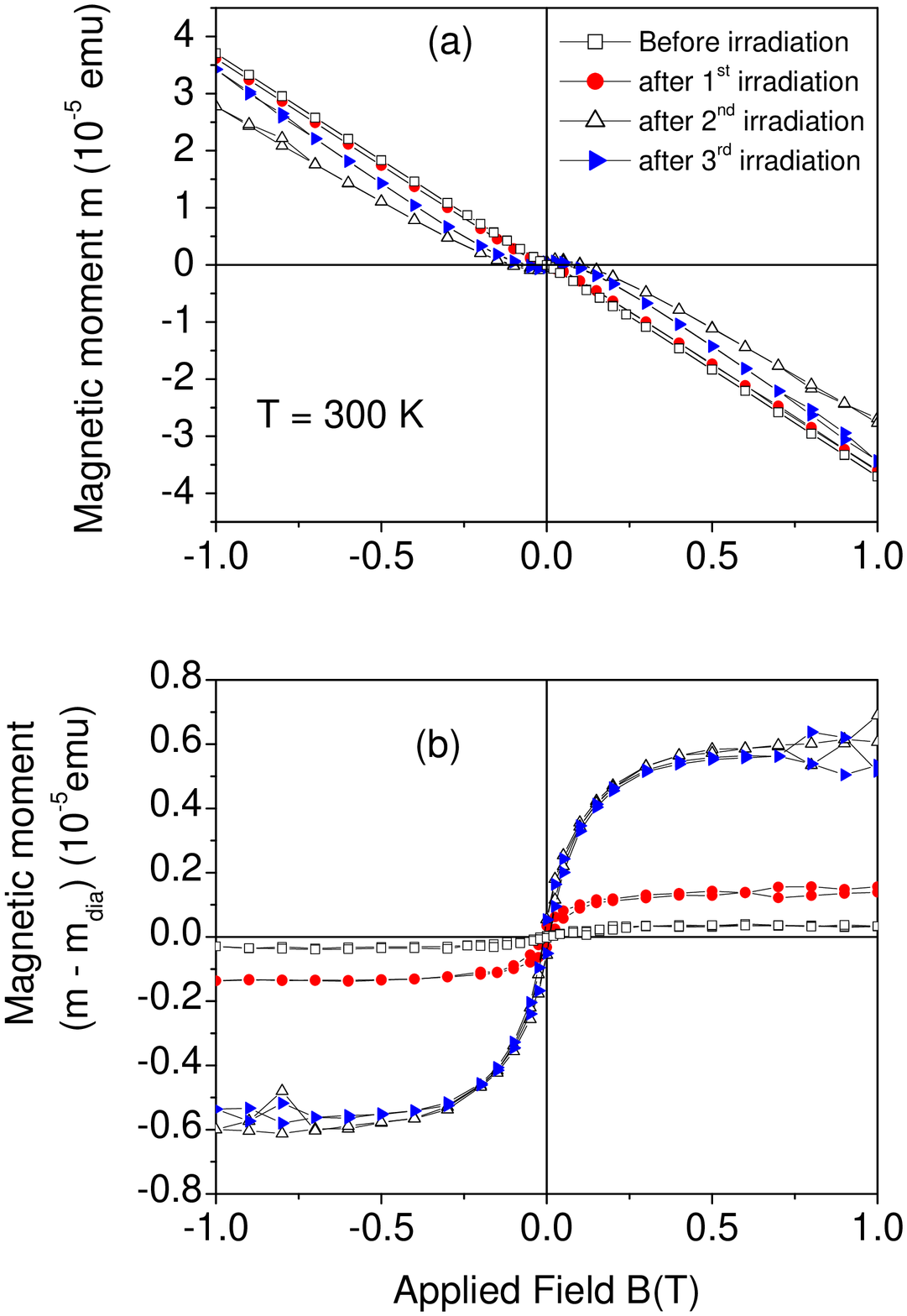}
\end{center}
\vspace{0.0cm} \caption{(a) Magnetic moment vs. applied field of the HOPG
sample AS171103/2 glued on a Si substrate before and after irradiation
(for details see text). (b) The same data as (a) but after substraction of
the diamagnetic background.} \label{be1711}
\end{figure}

\begin{figure}
\begin{center}
\includegraphics[width=\textwidth]{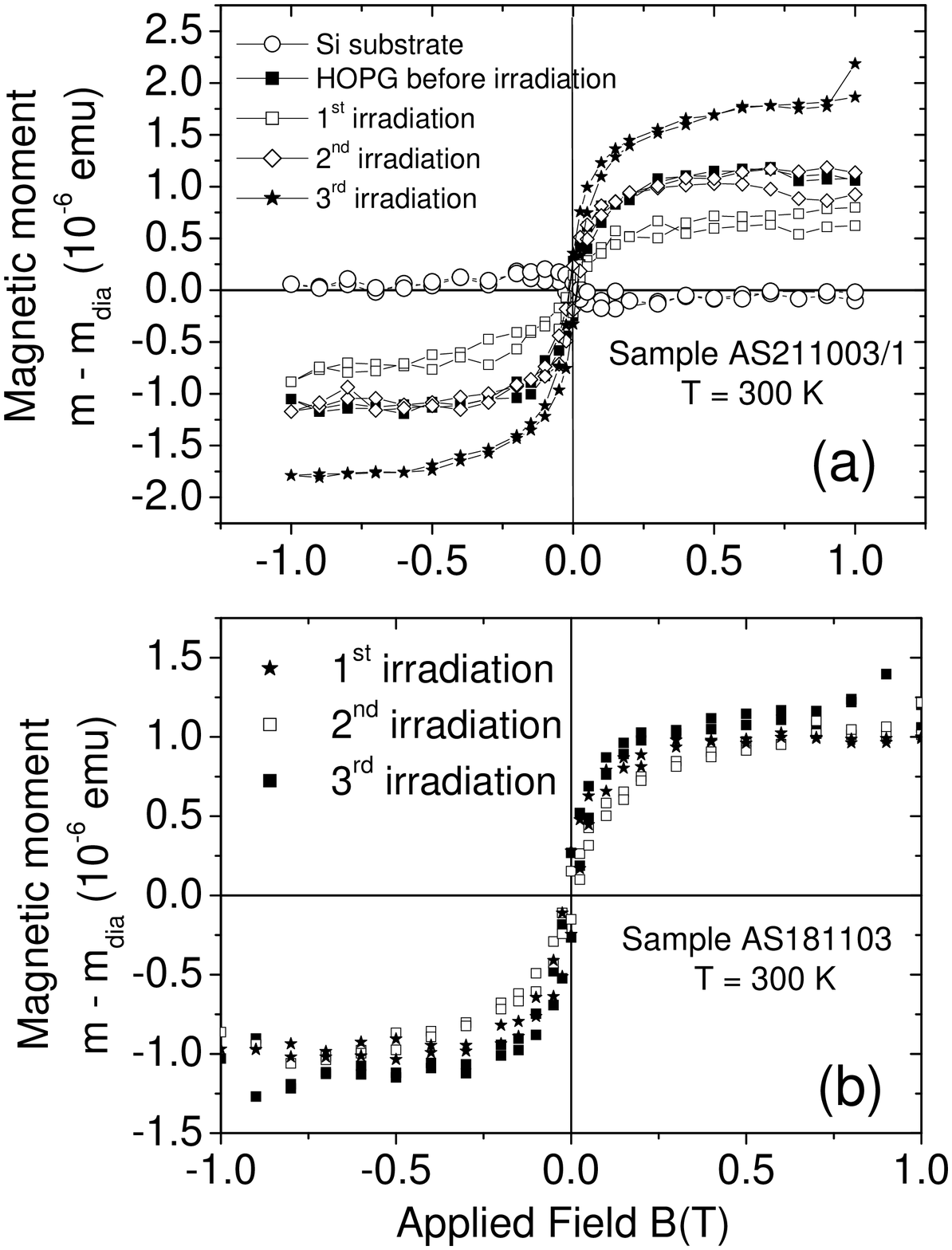}
\end{center}
\vspace{0.0cm} \caption{(a) Magnetic moment (minus the diamagnetic
background) vs. field of a typical Si substrate $(\circ)$ where all HOPG
samples were attached before and after all irradiation steps. The other
symbols correspond to sample AS211003/1 after different proton
irradiations (see text for details) with proton current $I \sim 55~$nA.
(b) The same for sample AS181103 after different proton irradiations with
a current $I \simeq 150~$nA (for details see text).} \label{2I}
\end{figure}

To study the influence of some of the irradiation parameters on
the magnetic response of HOPG samples, we have studied three HOPG
samples from the same batch. One sample  of mass $m = 1.19~$mg
(labelled AS171103/2) was irradiated with the proton microbeam as
follows. First irradiation consisted of $10^4$ spots of diameter
$\phi \simeq 1.8~\mu$m each with a charge of 0.85~nC (total charge
8.5~$\mu$C), fluence $= 0.34~$nC/$\mu$m$^2$, irradiated area:
$0.62 \times 0.62~$mm$^2$ and current $I \simeq 1.1$~nA. Second
irradiation produced the same amount of spots but with a diameter
of $\phi \simeq 2~\mu$m and 0.72~nC charge each (total charge
7.2~$\mu$C), fluence $= 0.24~$nC/$\mu$m$^2$ and $I \simeq 0.9$~nA.
The third and last irradiation  was identical to the first one.
The room temperature SQUID results of the sample before and after
substraction of the diamagnetic background are shown in
Fig.~\ref{be1711}. It is clearly seen that the irradiation
increases the magnetic signal of the sample. The last irradiation,
however, did not produce any significant change respect to the
last one. The remanent magnetization changed from $M_r (B = 0)
\simeq 3.3 \times 10^{-5}~$emu/g to $2.7 \times 10^{-4}~$emu/g and
$4.5 \times 10^{-4}~$emu/g for the first, virgin state and second
(or third) irradiation steps.

To check the influence of the irradiation current under broad irradiation
conditions we have irradiated two samples with the following
characteristics. Irradiation with a current $I = 54 \pm 1~$nA: sample
AS211003/1, mass $m = 1.33~$mg, first irradiation: 4 spots of $\phi =
0.8~$mm and $53.4~\mu$C each (total charge 214~$\mu$C); second
irradiation: 4 spots of $\phi = 0.8~$mm and $110~\mu$C each (total charge
440~$\mu$C); third irradiation: 4 spots of $\phi = 0.8~$mm and $54.5~\mu$C
each (total charge 218~$\mu$C). The SQUID results after substraction of
the diamagnetic background are shown in Fig.~\ref{2I}(a). We see that the
initial magnetic state of this sample (magnetic moment at saturation $m
\simeq 10^{-6}~$emu) decreased after the first irradiation. For the
subsequent irradiation steps the magnetic moment increased. The saturation
magnetization showed a decrease from its initial value of $M_r =(1.95 \pm
0.15) 10^{-4}~$emu/g to $M_r =(1.0 \pm 0.08) 10^{-4}~$emu/g after the
first irradiation. For the 2$^{\rm nd}$ and $3^{\rm rd}$ irradiation we
have $M_r =(1.56 \pm 0.15) 10^{-4}~$emu/g and $M_r =(2.6 \pm 0.15)
10^{-4}~$emu/g. The coercive field remains in the range of 150~Oe.

On the sample AS181103 of mass $m = 1.12~$mg, the following
irradiation steps at a proton current of $I = 150~$nA were
performed: (1) 2 spots of 0.8~mm diameter with 159~$\mu$C each
(total charge 318~$\mu$C); (2) 2 spots of 0.8~mm diameter with
160~$\mu$C each (total charge 320~$\mu$C); (3) 2 spots of 0.8~mm
diameter with 154~$\mu$C each (total charge 308~$\mu$C).
Figure~\ref{2I}(b) shows the SQUID results for this sample. Under
those irradiation conditions the sample did not show any
significant increase in the magnetic signal within experimental
error. Comparing this result with that of (a) we would conclude
that the proton current plays an important role in inducing the
magnetic ordering in HOPG samples.

\subsubsection{Magnetic spots and lines of micrometer size}
\label{spots}

\begin{figure}
\begin{center}
\includegraphics[width=\textwidth]{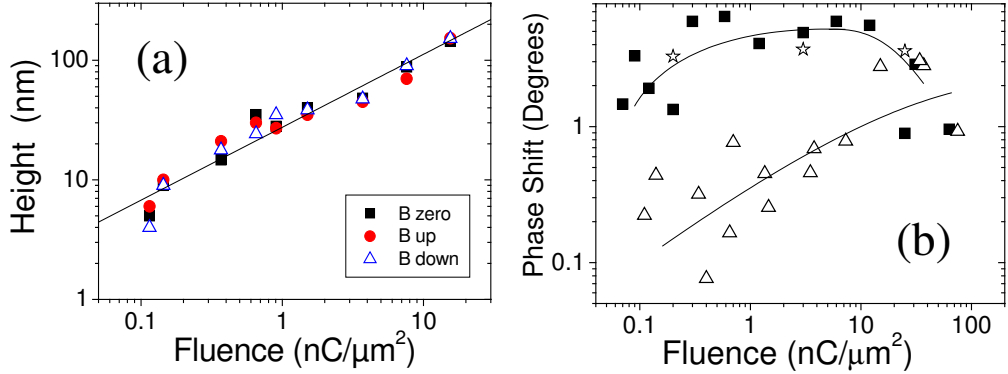}
\end{center}
\vspace{0.0cm} \caption{(a) Maximum height of swelling as a function of
the fluence, measured at micrometer small spots with an AFM. The three
symbols indicate the height measured before and after the application of a
magnetic field in $z$ (up) and $-z$ (down) direction. The measurements
were done without applied field. The line is the function $27.5
x^{0.61}~$nm with $x$ being the fluence in nC/$\mu$m$^2$. (b) Maximum
phase shift measured at micrometer small spots on two different HOPG
samples, before application of a magnetic field, as a function of the
irradiation fluence. The spot areas were $1~\mu$m$^2 (\blacksquare,\star)$
and $4~\mu$m$^2 (\vartriangle)$; the corresponding proton currents were
$171~$pA and $855 $~pA, respectively. The $(\star)$ symbols correspond to
the same sample measured again two months later. The lines are only a
guide to the eye.} \label{height}
\end{figure}

As mentioned in section~\ref{ic}, the LIPSION accelerator has the
possibility to irradiate the samples with a proton micro- or
nanobeam. We have used the microbeam to produce magnetic spots on
oriented graphite surfaces. As for the broad beam irradiation, the
microbeam was directed onto the HOPG surface parallel to the
$c-$axis of the sample without beam scanning (excepting line
scans, see below) leading to the formation of micron-sized spots
with enhanced defect density, as measured by micro-Raman, see
Fig.~\ref{raman}. In general, two spots separated by a distance of
$20~\mu$m were irradiated  with the same ion fluence and several
ion fluences  were used. For large enough fluences the swelling at
the spots can be directly observed with an optical microscope
\citep*{spemann04,pablopt}. The height of this swelling depends on
the irradiated fluence and on the mass of the ions. The dependence
of the maximum swelling height, measured by AFM, with the fluence
for proton irradiation is shown in Fig.~\ref{height}(a). With a
MFM one can measure the magnetic signal on the spots. The maximum
amplitude of the signal (maximum phase shift) as a function of the
fluence for two different proton currents is presented in
Fig.~\ref{height}(b). The units of the magnetic signal from the
MFM are ``degrees"; a relation of this phase shift of the tip
vibration to the force gradient can be found in
\citet*{lohau99,lohau00}. Examples for the magnetic moment
calculated from measurements of the phase shift in carbon samples
can be found in \citet*{hancar03,hanadd,hanap}.

We note that, whereas the swelling height increases systematically
with fluence, the maximum phase shift of the magnetic signal does
not, see Fig.~\ref{height}(b). This result indicates that: (1)
there is a negligible influence of the topo\-graphy onto the
magnetic signal and (2) the magnetic signal depends on the
implanted charge, proton current and probably also on the carbon
structure. The difference between the behaviors observed for two
different proton currents can be understood if we take into
account that the higher the current the higher would be the
internal temperature of the sample at the spot. Therefore,
annealing effects can be the origin for this difference. It is
worth to note that the signals could be reproduced after leaving
the samples in air and at room temperature for two months, see
Fig.~\ref{height}(b). This is not the case for much larger times,
see section \ref{ann}. We note also that at the higher fluence
range, the structure of the spots is very probably that of highly
disordered or even amorphous carbon. For both proton currents the
maximum magnetic signal decreases at the highest fluences, see
Fig.~\ref{height}(b).

\begin{figure}
\begin{center}
\includegraphics[width=10cm]{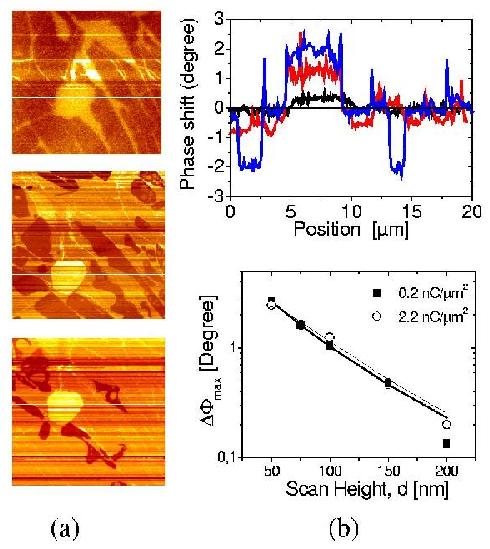}
\end{center}
\vspace{0.0cm} \caption{(a)Magnetic force gradient images $(20
\times 20~\mu$m$^2$) of a spot and its surroundings irradiated
with 0.115~nC/$\mu$m$^2$ (proton current $I =171~$pA). The images
were taken, from top to bottom, before field application, after
applying a field of $\sim 1~$kOe in the $+z$ direction parallel to
the $c$-axis, and in the $-z$ direction. The tip-to-sample
distance was 50~nm. (b) Top: The corresponding phase shift
obtained at the line scans (white straight lines in figures (a)).
The spot is located between $\sim 5~\mu$m and $\sim 10~\mu$m. The
bottom, upper and middle lines in this region correspond to
measurements before and after application of a field in $-z$ and
$+z$ direction. Adapted from \protect\citet*{han03}. Bottom:
Scanning height dependence of the maximum phase shift at proton
irradiated spots with fluences of 0.2~nC/$\mu$m$^2$ and
2.2~nC/$\mu$m$^2$. Solid lines are fits with the point probe
approximations. Adapted from \protect\citet*{hanap}.}
\label{figproton}
\end{figure}
\begin{figure}
\begin{center}
\includegraphics[width=12cm]{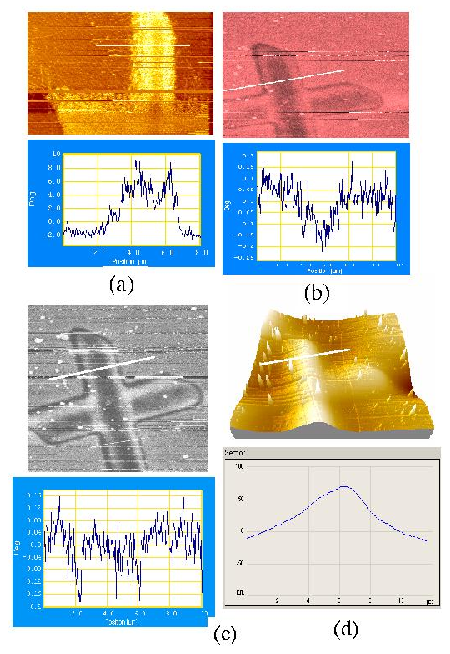}
\end{center}
\vspace{0.0cm} \caption{(a) Magnetic image (top, $8 \times
16~\mu$m$^2$) and the corresponding phase shift (bottom) obtained
at the line scan (white lines in all figures) for a cross produced
with a fluence of 0.5~nC/$\mu$m$^2$ and a proton current of $I =
120~$pA. The measurement was done before applying any magnetic
field and with a LM (low moment), well magnetized tip. (b) Similar
to (a) but for a cross produced with a fluence of 2~nC/$\mu$m$^2$
and after application of a field of 1~kOe normal to the surface.
The magnetic image has a size of $18 \times 18~\mu$m$^2$ and was
taken with a standard MESP tip. (c) The same cross as in (b) but
after application of a field of 2~kOe in the same direction as in
(b). The magnetic image has a size of $10 \times 20~\mu$m$^2$. (d)
Three dimensional image of the topography of the cross in (b,c)
and the corresponding line scan (bottom).} \label{cross}
\end{figure}

Figure~\ref{figproton} shows an example of a magnetic spot as measured
with the MFM before and after application of a magnetic field in two
directions. The irradiated region can be clearly recognized as the
``white" spot of the magnetic images in (a). A line scan through these
images indicates clear changes of the phase shift after application of a
magnetic field, see top figure in Fig.~\ref{figproton}(b). Usually the
magnetic images are obtained with a tip to sample distance of 50~nm. If we
increase this distance the phase shift amplitude decreases. This
dependence can be used to estimate, although with a relatively large
error, the order of magnitude of the magnetization at the spot surface.
From the data shown in the bottom figure of Fig.~\ref{figproton}(b) one is
able to estimate a magnetization of the order of 400~emu/g \citep*{hanap},
a very large magnetization of the order of soft ferromagnetic metals.

As a first attempt to write magnetically on a graphite surface we
have started with a simple cross. Figure \ref{cross}(a-c) shows
the magnetic images for two crosses made at different irradiation
fluences and measured with different tips and the topography for
one of them (d). The magnetic image changes after applying a
magnetic field. The results after magnetizing the cross show a
state with one or three magnetic domains. Micromagnetic
simulations with the appropriate parameters provide similar MFM
images \citep*{ziesem} and they may be used to obtain material
parameters (magnetic anisotropy, for example) that are not yet
possible to measure directly.

\subsubsection{Irradiation effects with alpha particles}
\label{alpha}

\begin{figure}
\begin{center}
\includegraphics[width=\textwidth]{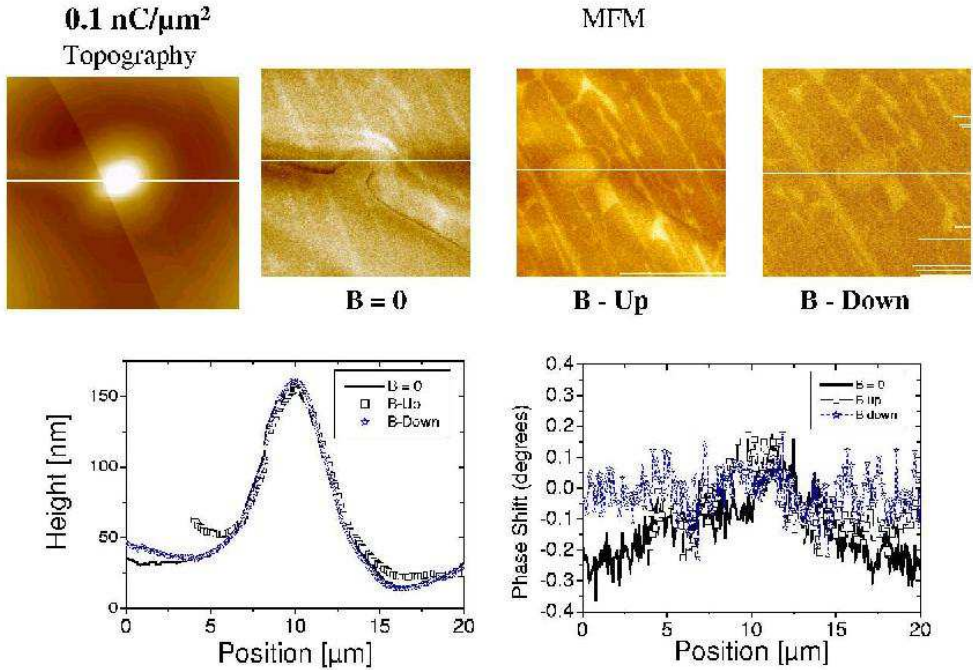}
\end{center}
\vspace{0.0cm} \caption{Micrometer spot produced after irradiation
of alpha particles  of 1.5 MeV on a HOPG surface with a fluence of
0.1~nC/$\mu$m$^2$. The upper pictures show the topography (left)
and the magnetic images before application of a magnetic field
(B=0), and after the application of a field in $+z$ (B-up) and in
$-z$ direction (B-down). The bottom pictures show the line scans
obtained at the position of the white lines in the upper images.}
\label{he1}
\end{figure}

With the LIPSION accelerator we have also the possibility to
irradiate the samples with a microbeam of 1.5 MeV alpha (He$^+$)
particles. The SRIM2003 simulation indicates that the penetration
depth of these particles in graphite should be 4~$\mu$m. For a
fluence of 0.1~nC/$\mu$m$^2$ the defect density at the surface
should be 1.4\% and the region at the end of ion range should be
rather amorphous. As for protons, we have produced pairs of spots
at different fluences and studied the magnetic signal at the spot
position with MFM. Figure \ref{he1} shows the topography and
magnetic images and the line scans of a spot produced with a
fluence of 0.1~nC/$\mu$m$^2$. Whereas the swelling produced by the
He$^+$ irradiation is remarkable, we recognize that the maximum
phase shift measured by the MFM at the spot and its surroundings,
see Fig.~\ref{he1}, is much smaller than those obtained at the
spots produced with protons, see Fig.~\ref{figproton}.
Nevertheless, the fact that we can measure small
but not zero signals, indicates that ion-, other than proton-,
irradiation in the carbon structure may also trigger magnetic
ordering. As we mentioned above, we do not know yet whether the
hydrogen already present in the sample before irradiation or in
the irradiation environment (rest gases in the chamber, for
example) plays any role and therefore we should take these results
with caution as a proof for a hydrogen-independent magnetic
ordering. Future work should try to measure  HOPG samples
irradiated with alpha particles at different fluences and with the
SQUID.

 \subsection{On thin films}
 \label{thin}

\begin{figure}
\begin{center}
\includegraphics[width=\textwidth]{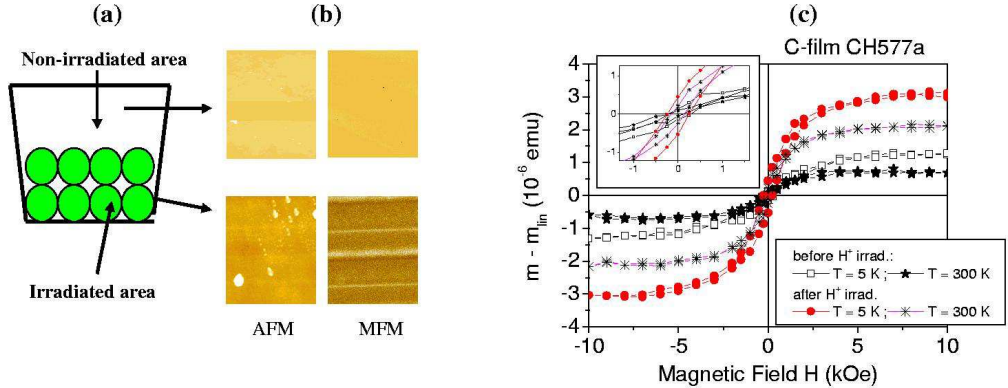}
\end{center}
\vspace{0.0cm} \caption{(a) Sketch of the disordered carbon film
CH577a produced by pulsed laser deposition with the irradiated (8
spots with 0.8~mm diameter and 150$~\mu$C total charge each) and
non-irradiated areas. (b) Topography (AFM) and magnetic (MFM)
images in both areas. (c) Magnetic moment of the film (after
substraction of the linear diamagnetic background) as a function
of magnetic field applied parallel to the main area. The symbols
$(\square,\star)$ indicate: before irradiation, measurements done
at $T = 5~$K and 300~K, and after proton irradiation,
$(\bullet,\ast)$ at 5~K and 300~K. The inset shows an enhanced
part around zero field. Adapted from \protect\citet*{hoh04}.}
\label{c-film}
\end{figure}

Taking into account the irradiation profile of protons in carbon
obtained from Monte Carlo simulations, see Fig.~\ref{protons}, one
would expect no or negligible effects on the magnetic properties
of samples, which thickness is much less than $\sim 40~\mu$m,
because most of the protons would go through the sample leaving a
small density of defects. We have checked this assumption
irradiating carbon and fullerene films with thickness below
$1~\mu$m. Figure~\ref{c-film} shows the MFM and SQUID results of a
carbon film produced by pulsed laser deposition (PLD) in a
hydrogen (H$_2$) atmosphere and deposited on a Si substrate. On
the film 8 spots of 0.8~mm diameter were irradiated (see sketch in
the figure). The MFM signal shows magnetic domains in the
irradiated region only, see Fig.~\ref{c-film}(b). The SQUID shows
a clear increase in the magnetic moment after irradiation, see
Fig.~\ref{c-film}(c). Because the results are taken on an
disordered sample, they also reveal that a graphite ordered
structure, at least in the mesoscopic level, is not necessary to
trigger magnetic order after irradiation.

\begin{figure}
\begin{center}
\includegraphics[width=\textwidth]{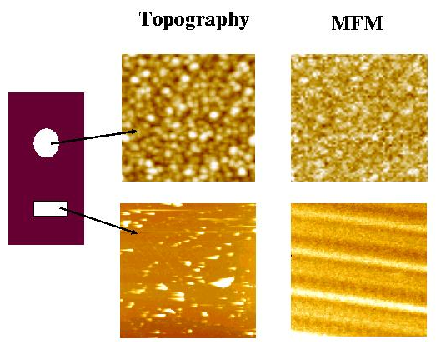}
\end{center}
\vspace{0.0cm} \caption{Left side sketch shows the fullerene film with the
two irradiated regions: the upper ellipse corresponds to the laser
illuminated region \protect\citep*{maka03} and the lower rectangle to the
region where 20 spots of 1.8~$\mu$m diameter each where irradiated with a
proton microbeam with a current of 500~pA and a fluence between 0.068 and
68~nC/$\mu$m$^2$. In contrast to the spot irradiation in HOPG (see
section~\ref{spots}) no change in the topography has been detected after
irradiation of the spots in the fullerene film. All images correspond to
an area of $5 \times 5~\mu$m$^2$.} \label{c60film}
\end{figure}

\citet*{maka03} studied the magnetism in photopolymerized fullerene films
by MFM. The magnetic force gradient measurements revealed that laser- and
electron-beam irradiation of fullerene films produces magnetic images
highly correlated to the topography. This correlation is expected since
the polymerization shrinks the film material at the surface within a
certain penetration depth, which depends on the radiation characteristics,
producing topographic clusters with magnetic order. An example of the AFM
and MFM images in a region illuminated in air with an energy of 2.6~eV and
intensity of 200 mW/cm$^2$ is shown in Fig.~\ref{c60film}. The high
correlation between of magnetic grain formation and topography is well
recognizable. In contrast to the laser illumination on C$_{60}$ film and
proton irradiation in HOPG, proton irradiation of spots of micrometer size
on the fullerene film do not produce appreciable changes in the
topography. The magnetic domains are however notable, see
Fig.~\ref{c60film}.

These results obtained after irradiation of the thin films are
unexpected and indicate either the hydrogen concentration in the
sample or surrounding before irradiation is relevant and/or the
defect concentration in the first micrometer from the sample
surface and produced by the proton beam is not negligible. Because
the thickness of the film is below $1~\mu$m  most of the protons
go through the material.

 \section{Annealing and Aging Effects}
 \label{ann}

We have measured with the SQUID some of the proton irradiated samples
months after the irradiation. Depending on the sample, the enhancement of
the magnetic order decreased or even vanished after more than 8 months
leaving them at room temperature and in air. However, the magnetic signal
remained for some of them even after one year. The reason for the
different behavior is not known yet. In this section we shall summarize
results that provide evidence for aging effects in the magnetic ordering,
as well as annealing effects at high temperatures, measured by SQUID or
MFM. Taking into account previous reports on the behavior and diffusion of
hydrogen in graphite \citep*{atsumi} as well as aging effects at room
temperature in the magnetization of the fullerene C$_{60}$H$_{24}$
\citep*{antonov02} one may expect to observe some time dependence in the
magnetic response at the irradiated surface by MFM (or in the bulk
magnetic moment by SQUID) if H is involved in the induced magnetism and a
diffusion takes place. Two different measurements were done to test the
possible influence of H diffusion on the magnetic signals. We measured
different irradiated spots on HOPG and irradiated amorphous carbon film
just after irradiation and after leaving the samples several months at
room temperature. Other samples we annealed at high temperatures in vacuum
or He atmosphere.

\begin{figure}
\begin{center}
\includegraphics[width=\textwidth]{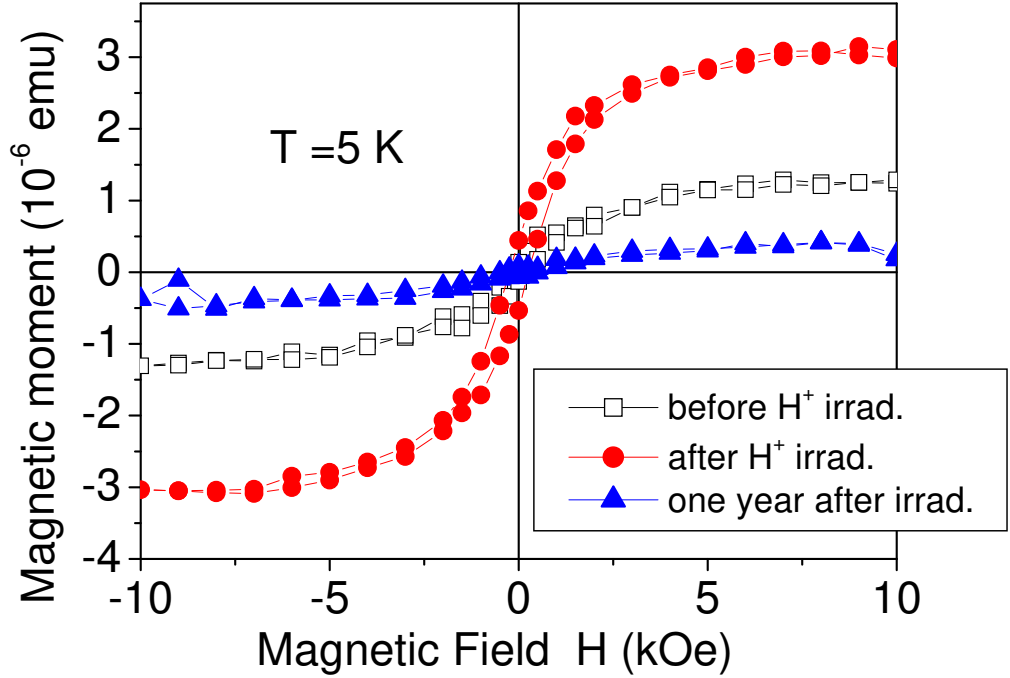}
\end{center}
\vspace{0.0cm} \caption{Magnetic moment (the diamagnetic
background was subtracted) as a function of applied field for the
disordered carbon film CH577a after preparation by PLD (before
irradiation $(\square)$), after irradiation ($\bullet$) and one
year later, leaving the sample at room temperature
($\blacktriangle$).} \label{cfilm1}
\end{figure}

Figure \ref{cfilm1} shows the magnetic moment of a disordered carbon
sample before, just after proton irradiation and after one year at room
temperature (after irradiation). It is clearly seen that the
ferromagnetic-like hysteresis vanishes after one year.

\begin{figure}
\includegraphics[width=\textwidth]{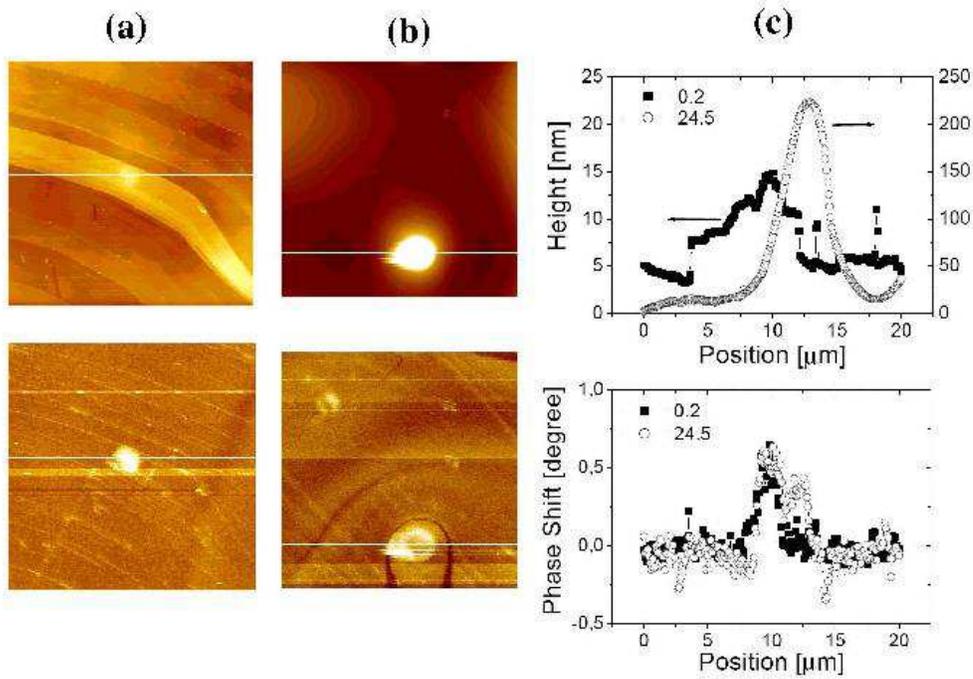}
 \caption{Topography
(top) and MFM (bottom) images ($20 \times 20~\mu$m$^2$) of sample
1 for two spots created at 0.2~nC/$\mu$m$^2$ (a) and 24.5
nC/$\mu$m$^2$ (b) and their surroundings measured one day after
irradiation. The tip-to-sample distance was 50 nm. (c)
Corresponding line scans (white lines in (a) and (b)) of
topography (top) and phase shift (bottom) images.}
\label{figspot1}
\end{figure}

Figures \ref{figspot1}(a) and (b) shows the topography (top) and
MFM (bottom) images of two spots and their surroundings created
with 0.2~nC/$\mu$m$^2$ and 24.5 nC/$\mu$m$^2$ on sample~1. The
measurement was done one day after proton irradiation. Line scans
of the topography (top) and MFM (bottom) images are shown in
Fig.~\ref{figspot1}(c). As shown in  this figure a clear
enhancement in both topography and MFM signals at the irradiated
region is clearly found. The enhancement of the topographic
swelling increases with fluence with our irradiation conditions.
For example, the peak height is $\sim 8$~nm ($\sim 225$~nm) for a
fluence of 0.2 nC/$\mu$m$^2$ (24.5 nC/$\mu$m$^2$). However, there
is no clear correlation between the increase in swelling height
measured by AFM and the maximum phase shift change of the MFM
signal at the spot within the fluence range used. As shown in
Fig.~\ref{figspot1}(c) bottom, a magnetic structure within the
spot produced with 24.5 nC/$\mu$m$^2$ appears, an indication for
the existence of magnetic domains within the area of the
irradiated spot.

\begin{figure}
\includegraphics[width=0.8\textwidth]{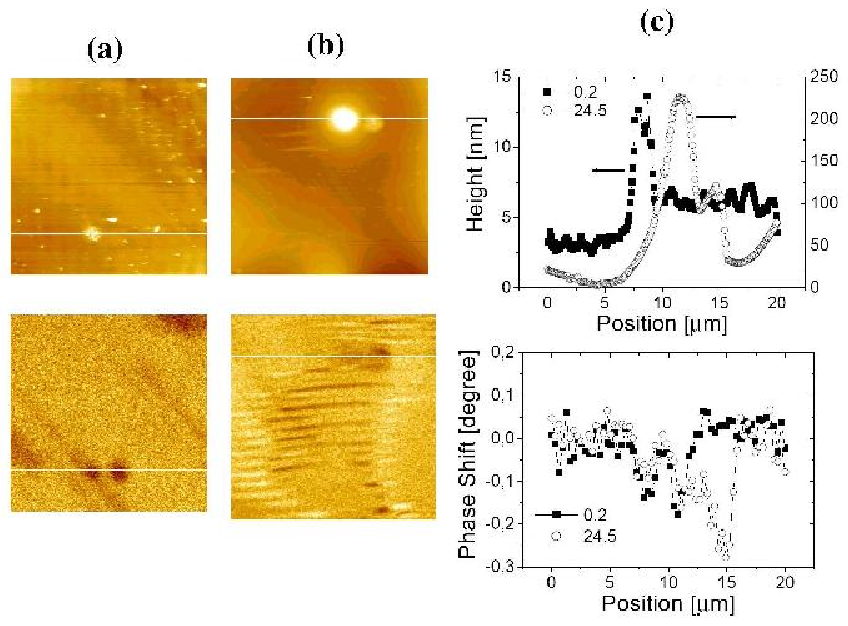}
\caption{(a) and (b) Topography (top) and MFM (bottom) images ($20
\times 20~\mu$m$^2$) of the same spots of
Fig.~\protect\ref{figspot1} and their surroundings measured after
8 months. The tip-to-sample distance was 50 nm. (c) Corresponding
line scans (white lines in (a) and (b)) of the topography (top)
and phase shift (bottom) images.} \label{figspot2}
\end{figure}

Figure \ref{figspot2} shows the same measurements as in
Fig.~\ref{figspot1}
 for the same two spots, but
obtained 8 months later. After this time we found clear changes in
the topography and in the phase shift of the MFM signal at and in
the surroundings of the original microspots, see
Fig.~\ref{figspot2}. There are three main changes to be noted: (1)
The sharply white spots in MFM images (see the bottom
Figs.~\ref{figspot1}(a) and (b)) on the irradiated area
 changed to black spots (see bottom  Figs.~\ref{figspot2} (a) and (b))
and the maximum value of phase shift changes from $\sim
+0.7^\circ$ to $\sim -0.2^\circ$. It means that the direction of
magnetization and its magnitude at the spots changed. (2) There
appear clear and localized MFM signals besides the original spot
signal (black spots in Figs.~\ref{figspot2}(a) and (b)). We note
also that these ``mirror" spots for both fluences are not clearly
related to the topographic swellings as can be seen by comparing
the line scan signals of topography and MFM of both spots (see
Fig.~\ref{figspot2}(c)). There is also a clear change in the
topography at the irradiated spots indicating a lattice
relaxation. (3) The MFM images reveal that after 8 months there is
a change not only at the spot position and its near neighborhood
but also at its surroundings. This change is clearly observable in
the vicinity of the spot made with a fluence 24.5 nC/$\mu$m$^2$
where a well discernible domain pattern appears, see bottom
Fig.~\ref{figspot2}(b).


Hydrogen desorption in graphite was and still is a matter of research.
Depending on the characteristics of the H-trapping this desorption may or
may not be accompanied by a lattice change. Therefore, it is not
straightforward to conclude that the aging effects we observed are due
only to the H-diffusion without any structural relaxation. Because we got
evidence for a structural relaxation, this can influence the effective
diffusion of H in the carbon matrix. Therefore, the effective activation
energies for H diffusion in our system are not necessarily the same as,
for example, those obtained by experimental methods, usually at high
temperatures, to study kinetics of diffusion of hydrogen in graphite
\citep*{atsumi}.

\begin{figure}
\includegraphics[width=\textwidth]{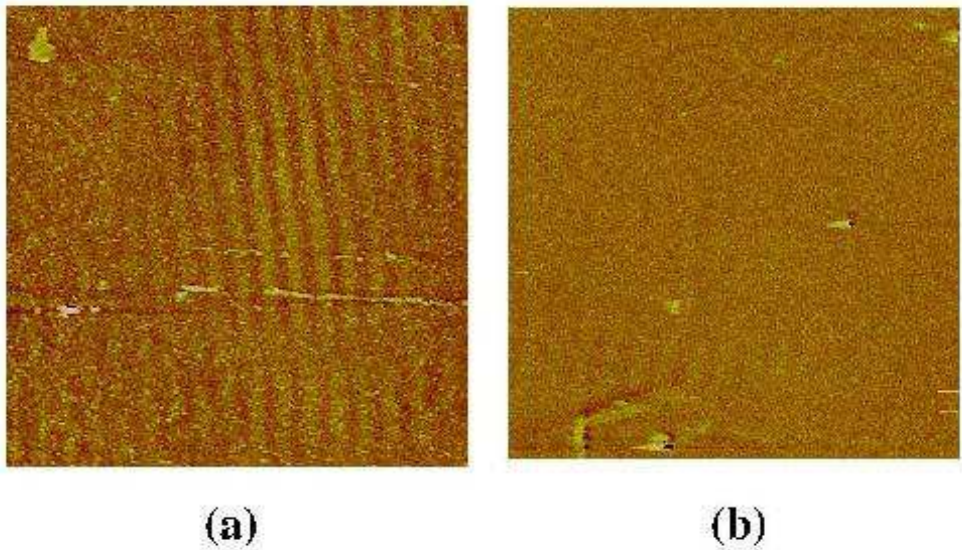}
\caption{The figures (a) and (b) show magnetic images of the same
crosses shown in Fig.~\protect\ref{cross}(a) and (c) measured
under similar conditions with similar MFM tips after annealing
them in vacuum at 1000~C for 2 hs. Whereas the crosses are not
anymore visible in MFM, domain structure is still detected in
(a).} \label{cran}
\end{figure}

Certainly, one does not necessarily wait one year to see effects
that may be related to hydrogen diffusion. It is quicker to anneal
an irradiated sample at high enough temperatures and check for its
influence. Figure \ref{cran} shows the MFM images obtained for the
crosses corresponding to those of Fig.~\protect\ref{cross}(a) and
(c) measured at identical conditions after annealing the sample
2~hs at 1000~C in vacuum. Whereas no signal is detected for the
cross in (b) ((c) in Fig.~\protect\ref{cross}) magnetic domains
are still identified after annealing in the surroundings of the
cross (a). Compare this observation to that obtained after aging
the spots at room temperature for several months shown in
Fig.~\ref{figspot2}.

\section{Conclusion and Open Issues}
\label{con}

In this chapter we have reviewed the main observations after proton
irradiation of carbon structures. The main results indicate that proton
irradiation can trigger magnetic ordering in graphite. The results after
proton irradiation leave no doubt that magnetic ordering exists in a
carbon structure without the influence of magnetic ions. Neither the total
amount of magnetic impurities is sufficient to account for the measured
magnetization nor the creation of magnetic spots in the micrometer range
with the proton micro-beam can be understood based on magnetic metal-ion
impurity concentration below 1~ppm as the PIXE results indicate. The
results obtained up today do indicate that different parameters and sample
states may play a role  in determining the strength of the induced
magnetic ordering. Broad irradiation (i.e. with a proton beam of large
diameter) appears to be less effective. The reason for this behavior will
be studied in the future, but we speculate that it may be due to the
produced defect density. Further experimental characterization (using the
broad spectrum of methods in magnetism research) and sample preparation
studies are necessary to understand and stabilize the magnetic ordering
found in carbon structures. The following issues should be clarified in
the near future:
\begin{enumerate}
\item The role of H-atoms, implanted by irradiation as well as
those already in the sample. \item The contribution to the
magnetic order from lattice defects produced by irradiation and
their possible influence as H-trapping centers. \item Dependence
of the magnetic order on the induced type of defect.  \item The
maximum achievable saturation magnetization in carbon structures.
\item The range of Curie temperature. \item Influence of the ion
current, fluence and energy of the irradiated particle on the
magnetic order. \item Influence of sample temperature. \item
Influence of the irradiation angle with respect to the
crystallographic $c$-axis of graphite. \item Influence of  ion
irradiation (other than proton) on the magnetism of carbon
structures. \item The effective magnetic moment of magnetic
impurities in graphite as well as in disordered carbon structures.
\end{enumerate}
An answer to part of these open questions will take several years
of research.

Summarizing briefly recently published theoretical
 work, which is correlated to the influence of hydrogen
 in carbon magnetism we
 note that:
 \begin{enumerate}
\item  Hydrogenated nanographite can have a
 spontaneous magnetization due to different numbers of mono- and
 dihydrogenated carbon atoms \citep*{kusakabe03}. Theoretical work
 using local-spin-density approximation calculates
the spin polarization of the graphite bands when different atoms as
hydrogen, fluor or oxygen are added on a graphene layer and predicts
magnetic ordering with fully or partially spin-polarized flat band, upon
the added atom \citep*{maru04}. Theoretical simulations indicate that a
magnetic ordered state should be easier to achieve with a mixture of
carbon-hydrogen bonds than fluorinated nanographite \citep*{maru04b}.
 Although the calculations were done
 attaching the hydrogen at the carbon atoms at the edges of a graphene
 layer, one may speculate that hydrogen can trigger the sp$^2$-sp$^3$ unbalance
 promoting  magnetic ordering in different carbon structures.
 \item
Recently published work indicates that hydrogen modifies substantially the
electronic structure of graphite around it. According to STM/AFM
measurements \citep*{rufi00} a single H-atom interacting with a graphite
surface modifies the electronic structure over a distance of 20 to 25
lattice constants. \item Muon spin rotation/relaxation experiments
\citep*{cha02} indicate that a positive muon  in graphite triggers a local
magnetic moment around it. Recently published theoretical work
\citep*{duplock04} supports this conclusion. It also indicates that
 upon the type of defect in a graphene layer, hydrogen may not
trigger any magnetic ordering. \item The magnetic moment and diffusion of
adatom defects in a graphite sheet were studied by \citet*{lehtinen03}.
The results of a full spin-polarized density functional theory indicate
that these defects may have a magnetic moment of about $0.5~\mu_B$. The
calculations of \citet*{lehtinen04} indicate that if an hydrogen
encounters an empty vacancy, then it compensates the dangling bond and the
magnetic moment of the vacancy vanishes. However, if a vacancy is
saturated by an hydrogen atom, a second hydrogen atom will bond to the
other side of the vacancy having a magnetic moment of $1.2~\mu_B$
localized on the dangling sp$^2$ bond. \item New theoretical and
experimental work \citep*{barbara04} suggests that hydrogen may play also
an important role on the magnetic ordering found in fullerenes.
\end{enumerate}

\noindent {\bf Acknowledgements}\\ This research is supported by
the Deutsche Forschungsgemeinschaft under DFG ES 86/11-1.

{\small


   \bibliographystyle{elsart-harv}
\bibliography{magnetic_carbon.bbl}
}

\end{document}